\documentclass[prd,aps,floatfix,nofootinbib]{revtex4}

\usepackage{amsmath,graphicx,color,epsfig}

\begin{document}

\title{$B \to K^{\ast}_0(1430) l^{+} l^{-}$ decays in supersymmetric theories}
\author{M. Jamil Aslam$^{1}$}
\author{Cai-Dian L\"{u}$^{2}$}
\author{Yu-Ming Wang$^{2}$}
\affiliation{$^{1}$Centre for Advanced Mathematics and Physics,
National University of Science and Technology, Rawalpindi, Pakistan}
\affiliation{$^{2}$Institute of High Energy Physics,  and
Theoretical Physics Center for Science Facilities,  CAS, P.O. Box
918(4), Beijing 100049, China}

\begin{abstract}
The weak decays of $B \to K^{\ast}_0(1430) l^{+} l^{-}$ ($l=\mu,
\tau$) are investigated in Minimal Supersymmetric Standard Model
(MSSM) and also in Supersymmetric (SUSY) SO(10) Grand Unified
Models. Neutral Higgs bosons are the point of main focus in MSSM
because they make quite a large contribution in exclusive $B \to
X_{s} l^{+} l^{-}$ decays at large $\tan{\beta}$ regions of
parameter space of SUSY models, as part of SUSY contributions is
proportional to $\tan^{3}{\beta}$. The analysis of decay rate,
forward-backward asymmetries and lepton polarization asymmetries in
$B \to K^{\ast}_0(1430) l^{+} l^{-}$ show that the values of these
physical observables are greatly modified by the effects of neutral
Higgs bosons. In SUSY SO(10) GUT model, the new physics contribution
comes from the operators which are induced by the neutral Higgs
boson penguins and also from the operators with chirality opposite
to that of the corresponding Standard Model (SM) operators. SUSY
SO(10) effects show up only in the decay $B \to K^{\ast}_0 \mu^{+}
\mu^{-}$ where the transverse lepton polarization asymmetries
deviate significantly from the SM value while the effects in the
decay rate, forward-backward asymmetries,  the longitudinal and
normal lepton polarization asymmetries are very mild. The transverse
lepton polarization asymmetry is almost zero in SM and in MSSM
model, whereas it can reach to $-0.3$ in SUSY SO(10) GUT model which
could be seen at the future colliders; hence this asymmetry
observable can be used to discriminate between different SUSY
models.
\end{abstract}

\pacs{13.30.Ce, 14.20.Mr, 11.30.Pb} \maketitle



\section{Introduction}

It is generally believed that the Standard Model (SM) is one of the
most successful theory of the second half of the last century as it
has passed all the experimental tests carried out for its
verifications. The only missing thing that is yet to be verified is
the Higgs boson mass, which we hope will be measured at the CERN
Large Hadron Collider (LHC) in next couple of years. To test the SM
indirectly, rare decays induced by flavor changing neutral currents
(FCNCs) $b \to s\, l^{+} l^{-}$ have become the main focus of the
studies since the CLEO measurement of the radiative decay $b \to
s\gamma$ \cite{CLEO1}. In the SM these decays are forbidden at tree
level and can only be induced    via loop diagrams. Hence, such
decays will provide useful information about the parameters of
Cabbibo-Kobayashi-Maskawa (CKM) matrix \cite{CKM 1, CKM 2} elements
as well as the various hadronic form factors. In literature, there
have been intensive studies on the exclusive decays $B \to P(V, A)
\, l^{+} l^{-}$ \cite{b to s in theory 16,b to s in theory 17,b to s
in theory 18,b to s in theory 19,b to s in theory
20,b to s in theory 21,jamil1} both in the SM and beyond, where the notions $%
P, V$ and $A$ denote the pseudoscalar, vector and axial vector mesons
respectively.

Despite all the success of SM no one can say that it is the ultimate
theory of nature as it has many open questions, such as gauge
hierarchy problem, origin of masses and Yukawa couplings, etc. It is
known that supersymmetry (SUSY) is not only one of the strongest
competitor of the SM but is also the most promising candidate of new
physics.  One direct way to search for SUSY is to discover SUSY
particles at high energy colliders, but unfortunately, so far no
SUSY particles have been found. Another way is to search for its
effects through indirect methods. The measurement of invariant mass
spectrum, forward-backward asymmetry and polarization asymmetries
are the suitable tools to probe new physics effects. For most of the
SUSY models, the SUSY contributions to an observable appear at loop
level due to the $\ R$-parity conservation. Therefore, it has been
realized for a long time that rare processes can be used as a good
probe for the searches of SUSY, since in these processes the
contributions of SUSY and SM arises at the same order in
perturbation theory \cite{Yan}.

In other SUSY models, Neutral Higgs Bosons (NHBs) could contribute
largely to the inclusive processes $B \to X_{s} l^{+} l^{-}$,
because part of the SUSY contributions is proportional to the
$\tan^{3}{\beta}$ \cite{Huang}. Subsequently, the physical
observables, like branching ratio and forward-backward asymmetry, in
the large $\mathrm{\tan{\beta}}$ region of parameter space in SUSY
models can be quite different from that in the SM. Motivated by the
fact, similar effects in exclusive $B \to K (K^{*})\,\ l^{+} l^{-}$
decay modes are also investigated \cite{Yan}, where the analysis of
decay rates, forward-backward asymmetries and polarization
asymmetries of final state lepton indicates the significant role of
NHBs. It is believed that the problem of neutrino oscillations can
not be explained in the SM. To this purpose, the SUSY SO(10) Grand
Unified Models (GUT)\cite{Gut} has been proposed in  literature. In
this model, there is a complex flavor non-diagonal down-type squark
mass matrix element of 2nd and 3rd generations which is of the order
one at the GUT scale. This can induce large flavor off-diagonal
coupling such as the coupling of gluino to the quark and squark
which belong to different generations. In general these couplings
are complex and may contribute to the process of FCNCs. The above
analysis of physical observables in $B \to K (K^{*})\,\ l^{+} l^{-}$
decay is extended in SUSY SO(10) GUT model where it has been shown
that the forward-backward asymmetries as well as the longitudinal
and transverse decay widths of the said decays are sensitive to
these NHBs effect in SUSY SO(10) GUT model which can be detected in
the future $B$ factories \cite{Li}.

In this paper, we will investigate the exclusive decay $\bar{B}_0
\to K^{\ast}_0(1430) l \bar{l}$ ( $l$ = $\mu $, $\tau $), where
$K^{\ast}_0(1430) $ is a scalar meson, both in the Minimal
Supersymmetric Standard Model (MSSM) as well as in the SUSY SO(10)
GUT model \cite{Gut}. We evaluate the branching ratios,
forward-backward asymmetries, lepton polarization asymmetries with
special emphasis on the effects of NHBs in MSSM. It is known that
different source of the vector current could manifest themselves in
different regions of phase space. For low value of momentum
transfer, the photonic penguin dominates, while the $Z$ penguin and
$W$ box become important towards high value of momentum transfer
\cite{Yan}. In order to search the region of momentum transfer with
large contributions from NHBs, the above decay in certain large
$\mathrm{{tan{\ \beta}}}$ region of parameter space has been
analyzed in SuperGravity (SUGRA) and M-theory inspired models
\cite{CSHuang}. We extend this analysis to the SUSY SO(10) GUT model
\cite{Yan}, where there are some primed counterparts of the usual SM
operators. For instance, the counterparts of usual operators in $\
B\rightarrow \ X_{s}\ \gamma $ decay are suppressed by $m_{s}/m_{b}$
and consequently negligible in the SM because they have opposite
chiralities. These operators are also suppressed in Minimal Flavor
Violating (MFV) models \cite{Bobeth,Wu}, however, in SUSY SO(10) GUT
model their effects can be significant. The reason is that the
flavor non-diagonal squark mass matrix elements are the free
parameters, some of which have significant effects in rare decays of
$B$ mesons \cite{Lunghi}.

The main job of investigating the semi-leptonic B meson decay is to
properly evaluate the hadronic matrix elements for $B \to
K^{\ast}_0(1430)$, namely the transition form factors, which are
governed by the non-perturbative QCD dynamics. Several methods exist
in the literature to deal with this problem,  among which the  QCD
sum rules approach  (QCDSR) \cite{QCDSR 1,QCDSR 2}   is a fully
relativistic approach and well rooted in quantum field theory.
However, short distance expansion fails in non-perturbative
condensate when applying the three-point sum rules to the
computations of form factors in the large momentum transfer or large
mass limit of heavy meson decays. As a marriage of standard QCDSR
technique and theory of hard exclusive process, the light cone QCD
sum rules (LCSR) \cite {LCQCDSR 1,LCQCDSR 2,LCQCDSR 3} cure the
problem of QCDSR applying to the large momentum transfer by
performing the operator product expansion (OPE) in terms of twist of
the relevant operators rather than their dimension \cite{braun
talk}. Therefore, the principal discrepancy between QCDSR and LCSR
consists in that non-perturbative vacuum condensates representing
the long-distance quark and gluon interactions in the short-distance
expansion are substituted by the light cone distribution amplitudes
(LCDAs) describing the distribution of longitudinal momentum carried
by the valence quarks of hadronic bound system in the expansion of
transverse-distance between partons in the infinite momentum frame.
An important advantage of LCSR is that it allows a systematic
inclusion of both hard scattering effects and the soft
contributions. Phenomenologically, LCSR has been widely applied to
investigate the semi-leptonic decays of heavy hadrons
\cite{Ball:1998tj, Khodjamirian:2000ds, Duplancic:2008ix,
Wang:2007fs}, radiative hadronic decays \cite{Ali:1993vd,
Aliev:1995wi,Wang:2008sm} and non-leptonic two body decays of $B$
meson \cite
{Khodjamirian:2000mi,Khodjamirian:2002pk,Khodjamirian:2003eq,Khodjamirian:2005wn}.

In our numerical analysis for $\bar{B}_0 \to K^{\ast}_0(1430)$
decays, we shall use the results of the form factors calculated by
LCSR approach in Ref. \cite{YuMing}, and the values of the relevant
Wilson coefficient for MSSM and SUSY SO(10) GUT models are borrowed
from Ref. \cite{Yan, Li}. The effects of SUSY contributions to the
decay rate and lepton polarization are also explored in this work.
Our results show that the decay rates are quite sensitive to the
NHBs contribution. The forward-backward asymmetry is zero in the SM
for these decays because of the absence of the scalar type coupling,
therefore any nonzero value of the forward-backward asymmetry will
give us indication of the new physics. It is known that the hadronic
uncertainties associated with the form factors and other input
parameters have negligible effects on the lepton polarization
asymmetries, therefore we have also studied these asymmetries in the
SUSY models mentioned above and found that the effects of NHBs are
quite significant in some regions of parameter space of SUSY.

The paper is organized as follows. In Sec. II, we present the
effective Hamiltonian for the semileptonic decay $B \to K^{\ast}_0
l^{+} l^{-}$. Section III contains the parameterizations and numbers
of the form factors for the said decay using the LCSR approach. In
Sec. IV we present the basic formulas of physical observables like
decay rates, forward-backward asymmetries and polarization
asymmetries of lepton in the above mentioned decay. Section V is
devoted to the numerical analysis of these observables and the brief
summary and concluding remarks are given in Sec. VI.

\section{Effective Hamiltonian}

By integrating out the heavy degrees of freedom in the full theory, the
general effective Hamiltonian for $\ b\rightarrow \ sl^{+}l^{-}$ in SUSY
SO(10) GUT model, can be written as \cite{Li}
\begin{eqnarray}
H_{eff} &=&-\frac{4 G_{F}}{\sqrt{2}}V_{tb}V_{ts}^{\ast }\bigg[{%
\sum\limits_{i=1}^{2}}C_{i}({\mu })O_{i}({\mu })+{\sum\limits_{i=3}^{10}(}%
C_{i}({\mu })O_{i}({\mu })+C_{i}^{\prime }({\mu })O_{i}^{\prime }({\mu }))
\nonumber \\
&&+\sum\limits_{i=1}^{8}{(}C_{Q_i}({\mu })Q_{i}({\mu })+C_{Q_i}^{\prime }({%
\mu })Q_{i}^{\prime }({\mu }))\bigg],  \label{effective haniltonian 1}
\end{eqnarray}
where $O_{i}({\mu })$ $(i=1,\ldots ,10)$ are the four-quark operators and $%
C_{i}({\mu })$ are the corresponding Wilson\ coefficients at the energy
scale ${\mu }$ \cite{Goto}. Using renormalization group equations to resum
the QCD corrections, Wilson coefficients are evaluated at the energy scale ${%
\mu =m}_{b}$. The theoretical uncertainties associated with the
renormalization scale can be substantially reduced when the
next-to-leading-logarithm corrections are included. The new
operators $Q_{i}({\mu })$ $(i=1,\ldots ,8)$ come from the NHBs
exchange diagrams, whose manifest forms and corresponding Wilson
coefficients can be found in \cite{Ewerth,Feng}. The primed
operators are the counterparts of the unprimed operators, which can
be obtained by flipping the chiralities in the corresponding
unprimed operators. It is believed that the effects of the
counterparts of usual chromo-magnetic and electromagnetic dipole
moment operators as well as semileptonic operators with opposite
chirality are suppressed by $m_{s}/m_{b} $ in the SM, but in SUSY
SO(10) GUTs their effect can be significant, since
$\delta^{dRR}_{23}$ can be as large as 0.5 \cite {Li,Gut}. Apart
from this, $\delta^{dRR}_{23}$ can induce new operators as the
counterparts of usual scalar operators in SUSY models due to NHB
penguins with gluino-down type squark propagator in the loop. It is
worth mentioning that these primed operators will appear only in
SUSY SO(10) GUT model and are absent in SM and MSSM \cite{Yan}.

The explicit expressions of the operators responsible for $B \to
K^{\ast}_0(1430) l^{+}l^{-}$ transition are given by
\begin{eqnarray}
O_{7} &=&\frac{e^{2}}{16\pi ^{2}}m_{b}\left( \bar{s}\sigma _{\mu \nu
}P_{R}b\right) F^{\mu \nu },\,\qquad O_{7}^{\prime }=\frac{e^{2}}{16\pi ^{2}}%
m_{b}\left( \bar{s}\sigma _{\mu \nu }P_{L}b\right) F^{\mu \nu }  \nonumber \\
O_{9} &=&\frac{e^{2}}{16\pi ^{2}}(\bar{s}\gamma _{\mu }P_{L}b)(\bar{l}\gamma
^{\mu }l),\,\qquad \ \ \ O_{9}^{\prime }=\frac{e^{2}}{16\pi ^{2}}(\bar{s}%
\gamma _{\mu }P_{R}b)(\bar{l}\gamma ^{\mu }l)  \nonumber \\
O_{10} &=&\frac{e^{2}}{16\pi ^{2}}(\bar{s}\gamma _{\mu }P_{L}b)(\bar{l}%
\gamma ^{\mu }\gamma _{5}l),\,\ \ \ \ \ \ O_{10}^{\prime }=\frac{e^{2}}{%
16\pi ^{2}}(\bar{s}\gamma _{\mu }P_{R}b)(\bar{l}\gamma ^{\mu }\gamma _{5}l)
\nonumber \\
Q_{1} &=&\frac{e^{2}}{16\pi ^{2}}(\bar{s}P_{R}b)(\bar{l}l),\qquad \qquad \ \
\ \ \ Q_{1}^{\prime }=\frac{e^{2}}{16\pi ^{2}}(\bar{s}P_{L}b)(\bar{l}l)
\nonumber \\
Q_{2} &=&\frac{e^{2}}{16\pi ^{2}}(\bar{s}P_{R}b)(\bar{l}\gamma _{5}l),\qquad
\ \ \ \ \ \ \ \ Q_{2}^{\prime }=\frac{e^{2}}{16\pi ^{2}}(\bar{s}P_{L}b)(\bar{%
l}\gamma _{5}l)  \label{relvent-operators}
\end{eqnarray}
with $P_{L,R}=\left( 1\pm \gamma _{5}\right) /2$. In terms of the
above Hamiltonian, the free quark decay amplitude for $b\rightarrow
s$ $l^{+}l^{-} $ can be derived as \cite{Huang}:
\begin{eqnarray}
\mathcal{M}(b &\rightarrow &sl^{+}l^{-})=-\frac{G_{F}\alpha}{\sqrt{2}\pi }%
V_{tb}V_{ts}^{\ast }\bigg\{C_{9}^{eff}(\bar{s}\gamma _{\mu }P_L b)(\bar{l}%
\gamma ^{\mu }l)+C_{10}(\bar{s}\gamma _{\mu }P_L b)(\bar{l}\gamma ^{\mu
}\gamma _{5}l)  \nonumber \\
&&-2m_{b}C_{7}^{eff}(\bar{s}i\sigma _{\mu \nu }\frac{q^{\nu }}{s}P_R b)(\bar{%
l}\gamma ^{\mu }l)+C_{Q_{1}}(\bar{s}P_R b)(\bar{l}l)+C_{Q_{2}}(\bar{s}P_R b)(%
\bar{l}\gamma _{5}l)+(C_{i}(m_{b})\leftrightarrow C_{i}^{\prime }(m_{b}))%
\bigg\}  ,  \label{quark-amplitude}
\end{eqnarray}
where $s=q^{2}$ and $q$ is the momentum transfer. The operator
$O_{10}$ can not be induced by the insertion of four-quark operators
because of the absence of the $Z$ boson in the effective theory.
Therefore, the Wilson coefficient $C_{10}$ does not renormalize
under QCD corrections and hence it is independent on the energy
scale. In addition to this, the above quark level decay amplitude
can
receive contributions from the matrix element of four-quark operators, $%
\sum_{i=1}^{6}\langle l^{+}l^{-}s|O_{i}|b\rangle $, which are usually
absorbed into the effective Wilson coefficient $C_{9}^{eff}(\mu )$, that one
can decompose into the following three parts \cite{b to s in theory 3,b to s
in theory 4,b to s in theory 5,b to s in theory 6,b to s in theory 7,b to s
in theory 8,b to s in theory 9}
\[
C_{9}^{eff}(\mu )=C_{9}(\mu )+Y_{SD}(z,s^{\prime })+Y_{LD}(z,s^{\prime }),
\]
where the parameters $z$ and $s^{\prime }$ are defined as $%
z=m_{c}/m_{b},\,\,\,s^{\prime }=q^{2}/m_{b}^{2}$. $Y_{SD}(z,s^{\prime })$
describes the short-distance contributions from four-quark operators far
away from the $c\bar{c}$ resonance regions, which can be calculated reliably
in the perturbative theory. The long-distance contributions $%
Y_{LD}(z,s^{\prime })$ from four-quark operators near the $c\bar{c}$
resonance cannot be calculated from first principles of QCD and are usually
parameterized in the form of a phenomenological Breit-Wigner formula making
use of the vacuum saturation approximation and quark-hadron duality. We will
neglect the long-distance contributions in this work because of the absence
of experimental data on $B\to J/\psi K^{\ast}_0(1430)$.  The manifest expressions for $%
Y_{SD}(z,s^{\prime })$ can be written as
\begin{eqnarray}
Y_{SD}(z,s^{\prime }) &=&h(z,s^{\prime })(3C_{1}(\mu )+C_{2}(\mu
)+3C_{3}(\mu )+C_{4}(\mu )+3C_{5}(\mu )+C_{6}(\mu ))  \nonumber \\
&&-\frac{1}{2}h(1,s^{\prime })(4C_{3}(\mu )+4C_{4}(\mu )+3C_{5}(\mu
)+C_{6}(\mu ))  \nonumber \\
&&-\frac{1}{2}h(0,s^{\prime })(C_{3}(\mu )+3C_{4}(\mu ))+{\frac{2}{9}}%
(3C_{3}(\mu )+C_{4}(\mu )+3C_{5}(\mu )+C_{6}(\mu )),
\end{eqnarray}
with
\begin{eqnarray}
h(z,s^{\prime }) &=&-{\frac{8}{9}}\mathrm{ln}z+{\frac{8}{27}}+{\frac{4}{9}}x-%
{\frac{2}{9}}(2+x)|1-x|^{1/2}\left\{
\begin{array}{l}
\ln \left| \frac{\sqrt{1-x}+1}{\sqrt{1-x}-1}\right| -i\pi \quad \mathrm{for}{%
{\ }x\equiv 4z^{2}/s^{\prime }<1} \\
2\arctan \frac{1}{\sqrt{x-1}}\qquad \mathrm{for}{{\ }x\equiv
4z^{2}/s^{\prime }>1}
\end{array}
\right. ,  \nonumber \\
h(0,s^{\prime }) &=&{\frac{8}{27}}-{\frac{8}{9}}\mathrm{ln}{\frac{m_{b}}{\mu
}}-{\frac{4}{9}}\mathrm{ln}s^{\prime }+{\frac{4}{9}}i\pi \,\,.
\end{eqnarray}

Apart from this, the non-factorizable effects \cite{b to s 1, b to s 2, b to
s 3,NF charm loop} from the charm loop can bring about further corrections
to the radiative $b\rightarrow s\gamma $ transition, which can be absorbed
into the effective Wilson coefficient $C_{7}^{eff}$. Specifically, the
Wilson coefficient $C^{eff}_{7}$ is given by \cite{c.q. geng 4}
\[
C_{7}^{eff}(\mu )=C_{7}(\mu )+C_{b\rightarrow s\gamma }(\mu ),
\]
with the absorptive part for the $b\rightarrow sc\bar{c}\rightarrow
s\gamma $ rescattering
\begin{eqnarray}
C_{b\rightarrow s\gamma }(\mu ) &=&i\alpha _{s}\bigg[{\frac{2}{9}}\eta
^{14/23}(G_{1}(x_{t})-0.1687)-0.03C_{2}(\mu )\bigg], \\
G_{1}(x) &=&{\frac{x(x^{2}-5x-2)}{8(x-1)^{3}}}+{\frac{3x^{2}\mathrm{ln}^{2}x%
}{4(x-1)^{4}}},
\end{eqnarray}
where $\eta =\alpha _{s}(m_{W})/\alpha _{s}(\mu )$, $%
x_{t}=m_{t}^{2}/m_{W}^{2}$. Here
we have dropped out the tiny contributions proportional to CKM sector $%
V_{ub}V_{us}^{\ast }$. In addition, $C_{7}^{\prime eff}(\mu )$ and $%
C_{9}^{\prime eff}(\mu )$ can be obtained by replacing the unprimed
Wilson coefficients with the corresponding prime ones in the above
formulae.

\section{Parameterizations of matrix elements and form factors in LCSR }

With the free quark decay amplitude available, we can proceed to
calculate the decay amplitudes for semi-leptonic decays of
$\bar{B}_0\to K^{\ast}_0(1430) l^{+} l^{-}$ at hadronic level, which
can be obtained by sandwiching the free quark amplitudes between the
initial and final meson states. Consequently, the following two
hadronic matrix elements
\begin{eqnarray}
\langle K^{\ast}_0(p)|\bar{s}\gamma _{\mu }\gamma _{5}b|B_{q^{\prime
}}(p+q)\rangle, \,\,\, \langle K^{\ast}_0(p)|\bar{s}\sigma _{\mu \nu
}\gamma _{5}q^{\nu }b|B_{q^{\prime }}(p+q)\rangle
\end{eqnarray}
need to be computed as can be observed from Eq. (\ref{effective
haniltonian 1}). The contributions from vector and tensor types of
transitions vanish due to parity conservations which is the property
of strong interactions. Generally, the above two matrix elements can
be parameterized in terms of a series of form factors as
\begin{eqnarray}
\langle K^{\ast}_0(p)|\bar{s}\gamma _{\mu }\gamma _{5}b|B_{q^{\prime
}}(p+q)\rangle &=&-i[f_{+}(q^{2})p_{\mu }+f_{-}(q^{2})q_{\mu }],
\label{axial form factor} \\
\langle K^{\ast}_0(p)|\bar{s}\sigma _{\mu \nu }\gamma _{5}q^{\nu
}b|B_{q^{\prime }}(p+q)\rangle
&=&-\frac{1}{m_{B}+m_{K^{\ast}_0}}\left[ \left( 2p+q\right) _{\mu
}q^{2}-\left( m_{B}^{2}-m_{S}^{2}\right) q_{\mu }\right] f_{T}\left(
q^{2}\right).  \label{tensor form factor}
\end{eqnarray}
At the large recoil region, these form factors satisfy the following
relations \cite {YuMing,Aliev:2007rq}
\begin{eqnarray}
f_{+}(q^2)={\frac{2 m_B }{m_B+m_{K^{\ast}_0}}} f_{T}(q^2), \qquad
f_{-}(q^2)=0, \label{relations of  form factor}
\end{eqnarray}
\begin{eqnarray}
f_{T}(q^2)=-{\frac{m_b -m_{q_2} }{m_{B} -m_{K^{\ast}_0}}}
f_{+}(q^2),
\end{eqnarray}
with the help of the equations of motion in the heavy quark limit
\cite {Gilani:2003hf}. Contracting Eqs. (\ref{axial form
factor}-\ref{tensor form factor}) with the four momentum $q^{\mu }$
on both side and making use of the equations of motion
\begin{eqnarray}
q^{\mu }(\bar{\psi}_{1}\gamma _{\mu }\psi _{2}) &=&(m_{2}-m_{1})\bar{\psi}%
_{1}\psi _{2}  \label{eq-motion1} \\
q^{\mu }(\bar{\psi}_{1}\gamma _{\mu }\gamma _{5}\psi _{2}) &=&-(m_{1}+m_{2})%
\bar{\psi}_{1}\gamma _{5 }\psi _{2}  \label{eq-motion}
\end{eqnarray}
we have
\begin{eqnarray}
\langle K^{\ast}_0(p)|\bar{s}\gamma _{5}b|B_{q^{\prime }}(p+q)\rangle &=&{\frac{%
-i}{m_{b}+m_{s}}}[f_{+}(q^{2})p \cdot q+f_{-}(q^{2})q^{2}]
\end{eqnarray}

To calculate the  non-perturbaive form factors, one has to rely on
some nonperturbative approaches. Considering the distribution
amplitudes up to twist-3, the form factors at small $q^2$ for
$\bar{B}_0\rightarrow K^{\ast}_0 l^{+}l^{-}$ have been calculated in
\cite{YuMing}  using the LCSR. The dependence of form factors
$f_{i}(q^2) (i=+,-,T)$ on  momentum transfer $q^2$ are parameterized
in either the single pole form
\begin{eqnarray}
\label{form} f_{i}(q^2)={f_i(0) \over 1-a_i q^2/m_{B_{0}}^{2}},
\label{single-pole model of form factors}
\end{eqnarray}
or the double-pole form
\begin{eqnarray}
\label{form2} f_{i}(q^2)={f_i(0) \over 1-a_i q^2/m_{B_{0}}^{2}+b_i
q^4/m_{B_{q_1}}^{4}}, \label{double-pole model of form factors}
\end{eqnarray}
in the whole kinematical region $0<q^2<(m_{B_{0}}-m_{K^{\ast}_0})^2$
while non-perturbative parameters $a_i$ and $b_i$ can be fixed by
the magnitudes of form factors corresponding to the small   momentum
transfer calculated in the LCSR approach. The  results for the
parameters $a_i $, $b_i$ accounting for the $q^2$ dependence of form
factors $f_{+}$, $f_{-}$ and $f_{T}$ are grouped in Table
\ref{di-fit B to Kstar0(1430)}.

\begin{table}[htb]
\caption{Numerical results for the parameters $f_i(0)$, $a_i$ and
$b_i$ involved in the double-pole fit of form factors
(\ref{double-pole model of form factors}) responsible for $\bar{B}_0
\to K^{\ast}_0(1430) l \bar{l}$ decay up to the twist-3 distribution
amplitudes of $K^{\ast}_0(1430)$ meson.} \label{di-fit B to
Kstar0(1430)}
\begin{tabular}{cccc}
\hline\hline
& $\hspace{2 cm} f_i(0)$ & $\hspace{2 cm} a_i$ & $\hspace{2 cm} b_i$ \\
\hline
$f_{+}$ & $\hspace{2 cm} 0.97^{+0.20}_{-0.20}$ & \hspace{2 cm} $%
0.86^{+0.19}_{-0.18}$ &  \\ \hline
$f_{-}$ & $\hspace{2 cm} 0.073^{+0.02}_{-0.02}$ & \hspace{2 cm} $%
2.50^{+0.44}_{-0.47}$ & \hspace{2 cm} $1.82^{+0.69}_{-0.76}$ \\ \hline
$f_{T}$ & $\hspace{2 cm} 0.60^{+0.14}_{-0.13}$ & \hspace{2 cm} $%
0.69^{+0.26}_{-0.27}$ &  \\ \hline\hline
\end{tabular}
\end{table}

\section{Formula for Physical Observables}

In this section, we are going to perform the calculations of some
interesting observables in phenomenology like the decay rates,
forward-backward asymmetry as well as the polarization asymmetries of final
state lepton. From Eq. (\ref{quark-amplitude}), it is straightforward to
obtain the decay amplitude for $\bar{B}_{0}\rightarrow K_{0}^{*}l^{+}l^{-}$
as
\begin{equation}
\mathcal{M}_{\bar{B}_{0}\rightarrow
K_{0}^{\ast}l^{+}l^{-}}=-\frac{G_{F}\alpha }{2\sqrt{2}\pi
}V_{tb}V_{ts}^{*}\left[ T_{\mu }^{1}(\bar{l}\gamma ^{\mu }l)+T_{\mu
}^{2}(\bar{l}\gamma ^{\mu }\gamma _{5}l)+T^{3}(\bar{l}l)\right] ,
\label{lambda-amplitude}
\end{equation}
where the functions $T_{\mu }^{1}$, $T_{\mu }^{2}$ and $T^{3}$ are
given by
\begin{equation}
T_{\mu }^{1}=i\left( C_{9}^{eff}-C_{9}^{\prime eff}\right)
f_{+}(q^{2})p_{\mu }+\frac{4im_{b}}{m_{B}+m_{K_{0}^{*}}}\left(
C_{7}^{eff}-C_{7}^{\prime eff}\right) f_{+}(q^{2})p_{\mu },
\label{(first-aux-function)}
\end{equation}
\begin{eqnarray}
T_{\mu }^{2} &=&i\left( C_{10}-C_{10}^{\prime }\right) \left(
f_{+}(q^{2})p_{\mu }+f_{-}(q^{2})q_{\mu }\right)
-\frac{i}{2m_{l}\left( m_{b}+m_{s}\right) }\left(
C_{Q_{2}}-C_{Q_{2}}^{\prime }\right) \left( f_{+}(q^{2})p\cdot
q+f_{-}(q^{2})q^{2}\right) q_{\mu },  \label{second-aux-function}
\end{eqnarray}
and
\begin{equation}
T^{3}=i\left( C_{Q_{1}}-C_{Q_{1}}^{\prime }\right) \frac{1}{m_{b}+m_{s}}%
\left( f_{+}(q^{2})p\cdot q+f_{-}(q^{2})q^{2}\right) .
\label{third-aux-function}
\end{equation}
It needs to point out that the terms proportional to $q_{\mu }$ in $T_{\mu
}^{1}$, namely $f_{-}(q^{2})$ does not contribute to the decay amplitude
with the help of the equation of motion for lepton fields. Besides, one can
also find that the above results can indeed reproduce that obtained in the
SM with $C_{i}^{\prime }=0$ and $T^{3}=0$.

\subsection{The differential decay rates and forward-backward asymmetry of
$\bar{B}_0\rightarrow K^{\ast}_0(1430) l^{+}l^{-}$}

The semi-leptonic decay $\bar{B}_0 \to K^{\ast}_0(1430) l^+ l^-$ is
induced by FCNCs. The differential decay width of $\bar{B}_0 \to
K^{\ast}_0(1430) l^+ l^-$ in the rest frame of $\bar{B}_0$ meson can
be written as \cite{PDG}
\begin{equation}
{d\Gamma(\bar{B}_0 \to K^{\ast}_0(1430) l^+ l^-) \over d q^2} ={1
\over (2 \pi)^3} {1 \over 32 m_{\bar{B}_0}} \int_{u_{min}}^{u_{max}}
|{\widetilde{M}}_{\bar{B}_0 \to K^{\ast}_0(1430) l^+ l^-}|^2 du,
\label{differential decay width}
\end{equation}
where $u=(p_{K^{\ast}_0(1430)}+p_{l^-})^2$ and
$q^2=(p_{l^+}+p_{l^-})^2$; $p_{K^{\ast}_0(1430)}$, $p_{l^+}$ and
$p_{l^-}$ are the four-momenta vectors of $K^{\ast}_0(1430)$, $l^+$
and $l^-$ respectively; $|{\widetilde{M}}_{\bar{B}_0 \to
K^{\ast}_0(1430) l^+ l^-}|^2$ is the squared decay amplitude after
integrating over the angle between the lepton $l^-$ and
$K^{\ast}_0(1430)$ meson. The upper and lower limits of $u$ are
given by
\begin{eqnarray}
u_{max}&=&(E^{\ast}_{K^{\ast}_0(1430)}+E^{\ast}_{l^-})^2-(\sqrt{E_{K^{\ast}_0(1430)}^{\ast
2}-m_{K^{\ast}_0(1430)}^2}-\sqrt{E_{l^-}^{\ast 2}-m_{l^-}^2})^2, \nonumber\\
u_{min}&=&(E^{\ast}_{K^{\ast}_0(1430)}+E^{\ast}_{l^-})^2-(\sqrt{E_{K^{\ast}_0(1430)}^{\ast
2}-m_{K^{\ast}_0(1430)}^2} +\sqrt{E_{l^-}^{\ast 2}-m_{l^-}^2})^2;
\end{eqnarray}
where the energies of $K^{\ast}_0(1430)$ and $l^-$ in the rest frame
of lepton pair $E^{\ast}_{K^{\ast}_0(1430)}$ and $E^{\ast}_{l^-}$
are   determined as
\begin{equation}
E^{\ast}_{K^{\ast}_0(1430)}= {m_{\bar{B}_0}^2
-m_{K^{\ast}_0(1430)}^2 -q^2 \over 2 \sqrt{q^2}}, \hspace {1 cm}
E^{\ast}_{l}={q^2\over 2\sqrt{q^2}}.
\end{equation}
Collecting everything together, one can write the general expression
of the differential decay rate for $ \bar{B}_0\rightarrow
K^{\ast}_0(1430)l^+ l^-$ as \cite{Chen:2007na}:
 \begin{eqnarray}
\frac{d\Gamma }{ds} &=&\frac{G_{F}^{2}\alpha ^{2}\left|
V_{tb}V_{ts}^{*}\right| ^{2}}{3072m_{B}^{3}\pi ^{5}s}\sqrt{1-\frac{4m_{l}^{2}%
}{s}}\sqrt{\lambda (m_{B}^{2},m_{K^{\ast}_0}^{2},s)}\times   \nonumber \\
&&\bigg\{\left| A\right| ^{2}\left( 2m_{l}^{2}+s\right) \lambda
+12sm_{l}^{2}\left( m_{B}^{2}-m_{K^{\ast}_0}^{2}-s\right) \left(
CB^{*}+C^{*}B\right) +12m_{l}^{2}s^{2}\left| C\right| ^{2}+6t\left|
D\right| ^{2}\left(
t-4m_{l}^{2}\right)   \nonumber \\
&&+\left| B\right| ^{2}\left( \left( 2m_{l}^{2}+s\right) \left(
m_{B}^{4}-2m_{B}^{2}m_{K^{\ast}_0}^{2}-2sm_{K^{\ast}_0}^{2}\right)
+\left( m_{K^{\ast}_0}^{2}-s\right) ^{2}+2m_{l}^{2}\left(
m_{K^{\ast}_0}^{4}+10tm_{K^{\ast}_0}^{2}+s^{2}\right) \right)
\bigg\},
  \label{drate}
\end{eqnarray}
where
\begin{equation}
\lambda =\lambda
(m_{B}^{2},m_{K^{\ast}_0}^{2},s)=m_{B}^{4}+m_{K^{\ast}_0}^{4}+s^{2}-2m_{B}^{2}m_{K^{\ast}_0}^{2}-2m_{K^{\ast}_0}^{2}s-2sm_{B}^{2}.
\label{function1}
\end{equation}
The auxiliary functions are defined as
 \begin{eqnarray}
A &=&i\left( C_{9}^{eff}-C_{9}^{\prime eff}\right) f_{+}(q^{2})+\frac{4im_{b}%
}{m_{B}+m_{K^{\ast}_0}}\left( C_{7}^{eff}-C_{7}^{\prime eff}\right)
f_{T}(q^{2})
\nonumber \\
B &=&i\left( C_{10}-C_{10}^{\prime }\right) f_{+}(q^{2})  \nonumber \\
C &=&i\left( C_{10}-C_{10}^{\prime }\right) f_{-}\left( q^{2}\right) +\frac{i%
}{2m_{e}\left( m_{b}+m_{s}\right) }\left( p\cdot
qf_{+}(q^{2})+q^{2}f_{T}(q^{2})\right) \left(
C_{Q_{2}}-C_{Q_{2}}^{\prime
}\right)   \nonumber \\
D &=&\frac{i}{m_{b}+m_{s}}\left( p\cdot
qf_{+}(q^{2})+q^{2}f_{T}(q^{2})\right) \left(
C_{Q_{1}}-C_{Q_{1}}^{\prime }\right)   \label{nauxfunction}
\end{eqnarray}

The forward-backward asymmetry for the decay modes $\bar{B}_0 \to
K^{\ast}_0(1430) l^+ l^-$ is exactly equal to zero in the SM
\cite{Belanger,Geng} due to the absence of scalar-type coupling
between the lepton pair, which serves as a valuable ground to test
the SM precisely as well as bound its extensions stringently. The
differential forward-backward asymmetry  of final state leptons in
different SUSY models can be written as
\begin{equation}
{\frac{dA_{FB}(q^{2})}{ds}}=\int_{0}^{1}d\cos \theta
{\frac{d^{2}\Gamma
(s,\cos \theta )}{dsd\cos \theta }}-\int_{-1}^{0}d\cos \theta {\frac{%
d^{2}\Gamma (s,\cos \theta )}{dsd\cos \theta }}
\end{equation}
and
\begin{equation}
A_{FB}(q^{2})={\frac{\int_{0}^{1}d\cos \theta {\frac{d^{2}\Gamma
(s,\cos \theta )}{dsd\cos \theta }}-\int_{-1}^{0}d\cos \theta
{\frac{d^{2}\Gamma
(s,\cos \theta )}{dsd\cos \theta }}}{\int_{0}^{1}d\cos \theta {\frac{%
d^{2}\Gamma (s,\cos \theta )}{dsd\cos \theta }}+\int_{-1}^{0}d\cos \theta {%
\frac{d^{2}\Gamma (s,\cos \theta )}{dsd\cos \theta }}}}.
\end{equation}
Now putting everything together, we have
\begin{equation}
A_{FB}(s)=(1/{\frac{d\Gamma }{ds}})\frac{\alpha ^{2}G_{F}^{2}\left|
V_{tb}V_{ts}^{*}\right| ^{2}\lambda (m_{B}^{2},m_{K_{0}^{*}}^{2},s)}{%
1024m_{B}^{3}\pi ^{5}}m_{l}(1-\frac{4m_{l}^{2}}{s})(AD^{*}+A^{*}D).
\label{FBasymmetry}
\end{equation}
It is clear from the expressions of decay rate and forward-backward
asymmetry that the contribution of the NHBs as well as that of the
SUSY SO(10) GUT model comes in through the auxiliary functions
defined in Eq.(\ref{nauxfunction}). Hence these SUSY effects
manifest themselves in the numerical results of these observables.

\subsection{Lepton Polarization asymmetries of $\bar{B}_0\rightarrow
K^{\ast}_0(1430) l^{+}l^{-}$}

In the rest frame of the lepton $l^{-}$, the unit vectors along
longitudinal, normal and transversal component of the $l^{-}$ can be
defined as \cite{Aliev UED}:
\begin{eqnarray}
s_{L}^{-\mu } &=&(0,\vec{e}_{L})=\left( 0,\frac{\vec{p}_{-}}{\left| \vec{p}%
_{-}\right| }\right) ,  \nonumber \\
s_{N}^{-\mu } &=&(0,\vec{e}_{N})=\left( 0,\frac{\vec{p}_{K^{\ast}_0
}\times \vec{p}_{-}}{\left| \vec{p}_{K^{\ast}_0 }\times
\vec{p}_{-}\right| }\right) ,
\label{p-vectors} \\
s_{T}^{-\mu } &=&(0,\vec{e}_{T})=\left( 0,\vec{e}_{N}\times \vec{e}%
_{L}\right) ,  \nonumber
\end{eqnarray}
where $\vec{p}_{-}$ and $\vec{p}_{K_{0}^{*}}$ are the three-momenta
of the lepton $l^{-}$ and $K_{0}^{*}(1430)$ meson respectively in
the center mass (CM) frame of $l^{+}l^{-}$ system. Lorentz
transformation is used to boost the longitudinal component of the
lepton polarization to the CM frame of the lepton pair as
\begin{equation}
\left( s_{L}^{-\mu }\right) _{CM}=\left( \frac{|\vec{p}_{-}|}{m_{l}},\frac{%
E_{l}\vec{p}_{-}}{m_{l}\left| \vec{p}_{-}\right| }\right)
\label{bossted component}
\end{equation}
where $E_{l}$ and $m_{l}$ are the energy and mass of the lepton. The
normal and transverse components remain unchanged under the Lorentz
boost.

The longitudinal ($P_{L}$), normal ($P_{N}$) and transverse
($P_{T}$) polarizations of lepton can be defined as:
\begin{equation}
P_{i}^{(\mp )}(s)=\frac{\frac{d\Gamma }{ds}(\vec{\xi}^{\mp }=\vec{e}^{\mp })-%
\frac{d\Gamma }{ds}(\vec{\xi}^{\mp }=-\vec{e}^{\mp })}{\frac{d\Gamma }{ds}(%
\vec{\xi}^{\mp }=\vec{e}^{\mp })+\frac{d\Gamma }{ds}(\vec{\xi}^{\mp }=-\vec{e%
}^{\mp })}  \label{polarization-defination}
\end{equation}
where $i=L,\;N,\;T$ and $\vec{\xi}^{\mp }$ is the spin direction
along the leptons $l^{\mp }$. The differential decay rate for
polarized lepton $l^{\mp }$ in $\bar{B}_{0}\rightarrow
K_{0}^{*}(1430)l^{+}l^{-}$ decay along any
spin direction $\vec{\xi}^{\mp }$ is related to the unpolarized decay rate (%
\ref{differential decay width}) with the following relation
\begin{equation}
\frac{d\Gamma (\vec{\xi}^{\mp })}{ds}=\frac{1}{2}\left( \frac{d\Gamma }{ds}%
\right) [1+(P_{L}^{\mp }\vec{e}_{L}^{\mp }+P_{N}^{\mp
}\vec{e}_{N}^{\mp }+P_{T}^{\mp }\vec{e}_{T}^{\mp })\cdot
\vec{\xi}^{\mp }]. \label{polarized-decay}
\end{equation}
We can achieve the expressions of longitudinal, normal and
transverse polarizations for $\bar{B}_{0}\rightarrow
K_{0}^{*}(1430)l^{+}l^{-}$ decays as collected below. The
longitudinal lepton polarization can be written as
\begin{eqnarray}
P_{L}(s)=(1/{\frac{d\Gamma }{ds}})\frac{\alpha ^{2}G_{F}^{2}\left|
V_{tb}V_{ts}^{*}\right| ^{2}\lambda ^{3/2}(m_{B}^{2},m_{K_{0}^{*}}^{2},s)}{%
3072m_{B}^{3}\pi ^{5}}(1-\frac{4m_{l}^{2}}{s}) (AB^{*}+A^{*}B).
\label{expression-LP}
\end{eqnarray}
Similarly, the normal lepton polarization is
\begin{eqnarray}
P_{N}(s) &=&(1/{\frac{d\Gamma }{ds}})\frac{\alpha
^{2}G_{F}^{2}\left|
V_{tb}V_{ts}^{*}\right| ^{2}m_{l}}{4096m_{B}^{3}\pi ^{4}\sqrt{s}}\sqrt{1-%
\frac{4m_{l}^{2}}{s}} \bigg[(m_{B}^{2}-m_{K_{0}^{*}}^{2}+s)(
A^{*}B+AB^{*})-2s( A^{*}C+AC^{*})\bigg], \label{expression-PN}
\end{eqnarray}
and the transverse one is given by
\begin{eqnarray}
P_{T}(s) &=&(1/{\frac{d\Gamma }{ds}})\frac{-i\alpha
^{2}G_{F}^{2}\left|
V_{tb}V_{ts}^{*}\right| ^{2}\lambda ^{1/2}(m_{B}^{2},m_{K_{0}^{*}}^{2},s)}{%
4096 m_{B}^{3}\pi
^{4}}(1-\frac{4m_{l}^{2}}{s})(m_{B}^{2}-m_{S}^{2}+s) \bigg[
(A^{*}D-AD^{*})+2m_{l}(B^{*}C-BC^{*})\bigg]. \nonumber \\
\label{expression-TP}
\end{eqnarray}
The ${\frac{d\Gamma }{ds}}$ appearing in the above equation is the
one given in Eq. (\ref{drate}) and $\lambda
(m_{B}^{2},m_{K^{\ast}_0}^{2},s)$ is the same as that defined in Eq.
(\ref{function1}).

\section{Numerical Analysis}

In this section, we would like to present the numerical analysis of
decay rates, forward-backward asymmetries and polarization
asymmetries. The numerical values of Wilson coefficients and other
input parameters used in our analysis are borrowed from Ref.
\cite{Yan, Li, YuMing} and collected in Tables II, III and IV.
\begin{table}[tbh]
\caption{{}Values of input parameters used in our numerical analysis}
\label{Input parameters}
\begin{tabular}{cc}
\hline\hline $G_{F}=1.166\times 10^{-5}$ GeV$^{-2}$ & $\left|
V_{ts}\right|
=41.61_{-0.80}^{+0.10}\times 10^{-3}$ \\
$\left| {V_{tb}}\right| =0.9991$ & $m_{b}=\left( 4.68\pm 0.03\right) $ GeV
\\
{$m_{c}\left( m_{c}\right) =1.275_{-0.015}^{+0.015}$ GeV} & $m_{s}\left( 1%
\text{ GeV}\right) =\left( 142\pm 28\right) $ MeV \\ \hline
$m_{B^0}=5.28$ GeV & $m_{K_{0}^{*}}=1.43$ GeV \\ \hline \hline
\end{tabular}
\end{table}
\begin{table}[tbh]
\caption{{}Wilson Coefficients in SM\ and different SUSY models but
without NHBs contributions. The primed Wilson coefficients
corresponds to the operators which are opposite in helicities from
those of the SM operators.}
\begin{tabular}{ccccccc}
\hline
Wilson Coefficients & $C_{7}^{eff}$ & $C_{7}^{\prime eff}$ & $C_{9}$ & $%
C_{9}^{\prime }$ & $C_{10}$ & $C_{10}^{\prime }$ \\ \hline SM &
$-0.313$ & $0$ & $4.334$ & $0$ & $-4.669$ & $0$ \\ \hline SUSY I or
II& $+0.3756$ & $0$ & $4.7674$ & $0$ & $-3.7354$ & $0$ \\ \hline
SUSY III & $-0.3756$ & $0$ & $4.7674$ & $0$ & $-3.7354$ & $0$ \\
\hline\hline
SUSY SO(10) $\left( A_{0}=-1000\right) $ & $-0.219+0i$ & $0.039-0.038i$ & $%
4.275+0i$ & $0.011+0.0721i$ & $-4.732-0i$ & $-0.075-0.670i$ \\ \hline\hline
\end{tabular}
\end{table}
\begin{table}[tbh]
\caption{{}Wilson coefficient corresponding to NHBs contributions.
SUSY I corresponds to the regions where SUSY can destructively
contribute and can change the sign of $C_{7}$, but without
contribution of NHBs, SUSY II refers to the region where $\tan \beta
$ is large and the masses of the superpartners are relatively small.
SUSY III corresponds to the regions where $\tan \beta $ is large and
the masses of superpartners are relatively large. The primed Wilson
coefficients are   for the primed operators in
eq.(\ref{relvent-operators}) from NHBs contribution in SUSY SO(10)
GUT model.
 The values in
the parentheses  are for the $\tau $ case.} \label{NHB}
\begin{tabular}{ccccc}
\hline
Wilson Coefficients & $C_{Q_{1}}$ & $C_{Q_{1}}^{\prime }$ & $C_{Q_{2}}$ & $%
C_{Q_{2}}^{\prime }$ \\ \hline SM & $0$ & $0$ & $0$ & $0$ \\ \hline
SUSY I & $0$ & $0$ & 0 & $0$ \\ \hline SUSY II & $6.5\left(
16.5\right) $ & $0$ & $-6.5\left( -16.5\right) $ & $0$
\\ \hline\hline
SUSY III & $1.2\left( 4.5\right) $ & $0$ & $-1.2\left( -4.5\right) $ & $0$ \\
\hline\hline
SUSY SO(10) $\left( A_{0}=-1000\right) $ & $
\begin{array}{c}
0.106+0i \\
\left( 1.775+0.002i\right)
\end{array}
$ & $
\begin{array}{c}
-0.247+0.242i \\
\left( -4.148+4.074i\right)
\end{array}
$ & $
\begin{array}{c}
-0.107+0i \\
\left( -1.797-0.002i\right)
\end{array}
$ & $
\begin{array}{c}
-0.250+0.246i \\
\left( -4.202+4.128i\right)
\end{array}
$ \\ \hline\hline
\end{tabular}
\end{table}
In the subsequent analysis, we will focus on the parameter space of large $%
\tan \beta $, where the NHBs effects are significant owing to the fact that
the Wilson coefficients corresponding to NHBs are proportional to $%
(m_{b}m_{l}/m_{h}) \tan ^{3} \beta $ $(h=h^{0}$, $A^{0})$. Here, one $\tan
\beta $ comes from the chargino-up-type squark loop and $\tan ^{2}\beta $
comes from the exchange of the NHBs. At large value of $\tan \beta $ the $%
C_{Q_{i}}^{(^{\prime })}$ compete with $C_{i}^{(^{\prime })}$ and
can overwhelm $C_{i}^{(^{\prime })}$ in some region as can be seen
from the Tables III and IV \cite{Huang}. SUSY I corresponds to the
regions where SUSY can destructively contribute and can change the
sign of $C_{7}$, but without contribution of NHBs, SUSY II refers to
the region where $\tan \beta $ is large and the masses of the
superpartners are relatively small. SUSY III corresponds to the
regions where $\tan \beta $ is large and the masses of superpartners
are relatively large. The primed Wilson coefficients are   for the
primed operators in Eq.(\ref{relvent-operators}) from NHBs
contribution in SUSY SO(10) GUT model. As the NHBs are proportional
to the lepton mass, the values shown in the table are for $\mu $
case and $\tau $ case (the values in parentheses of table IV). Apart
from the large $\tan \beta $ limit, the other two conditions
responsible for the large contributions from NHBs are: (i) the mass
values of the lighter chargino and lighter stop should not be too
large; (ii) the mass splitting of charginos and stops should be
large, which also indicate large mixing between stop sector and
chargino sector \cite{Yan}. Once these conditions are satisfied, the process $%
B\rightarrow X_{s}\gamma $ will not only impose constraints on
$C_{7}$ but it also puts very stringent constraint on the possible
new physics. It is well known that the SUSY contribution is
sensitive to the sign of the Higgs mass term and SUSY contributes
destructively when the sign of this term becomes minus. It is
pointed out in literature \cite{Yan} that there exist considerable
regions of SUSY parameter space in which NHBs can largely contribute
to the process $b\rightarrow sl^{+}l^{-}$ due to change of the
sign of $C_{7}$ from positive to negative, while the constraint on $%
b\rightarrow s\gamma $ is respected. Also, when the masses of SUSY
particles are relatively large, say about $450$ GeV, there exist
significant regions in the parameter space of SUSY models in which
NHBs could contribute largely. However, in these cases $C_{7}$ does
not change its sign, because contributions of charged Higgs and
charginos cancel each other. We hope that these scalar mode decays
of $B$ mesons can be used to distinguish between these two regions
of SUSY.

\begin{figure}[tbp]
\begin{center}
\begin{tabular}{ccc}
\vspace{-2cm} \includegraphics[scale=0.6]{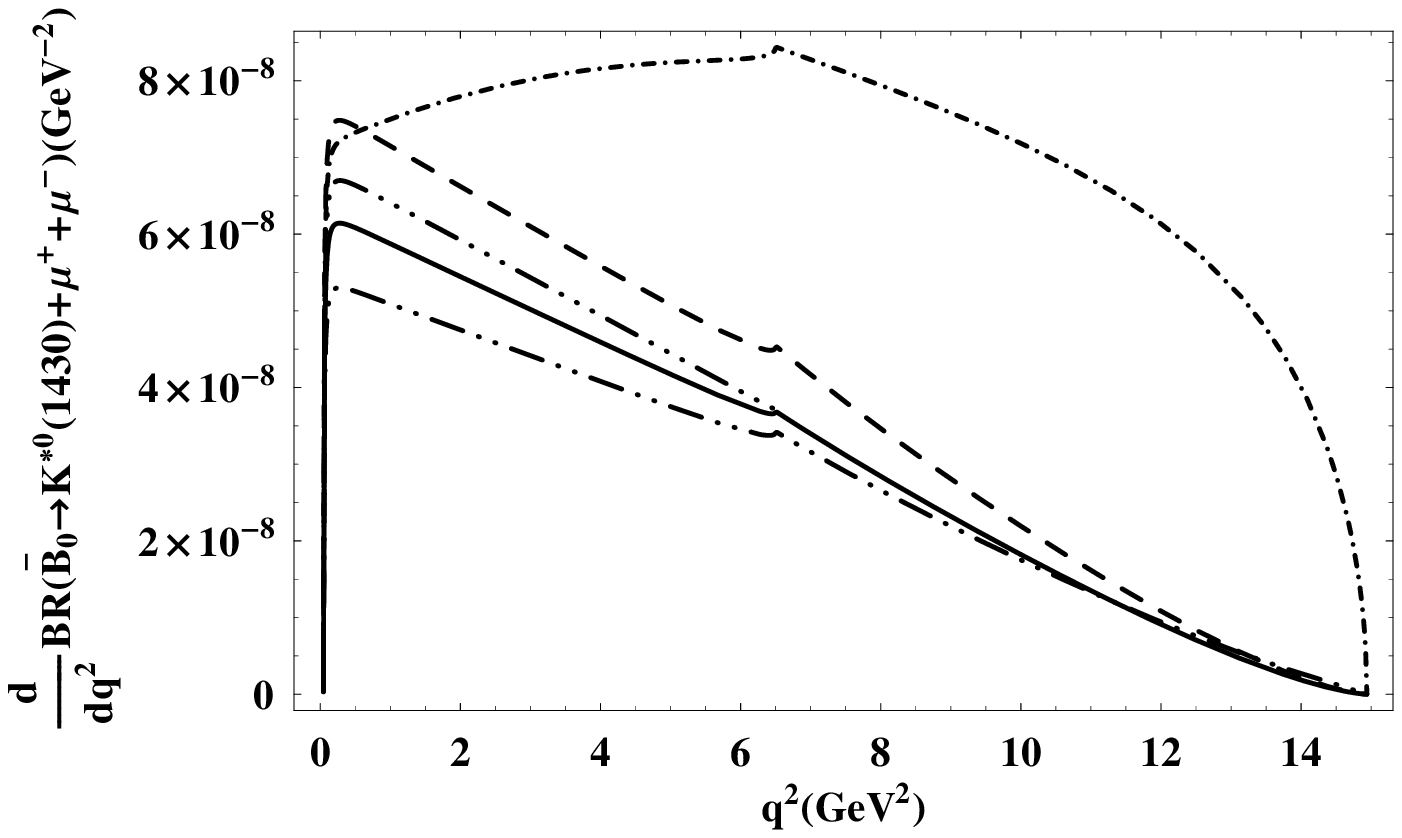} %
\includegraphics[scale=0.6]{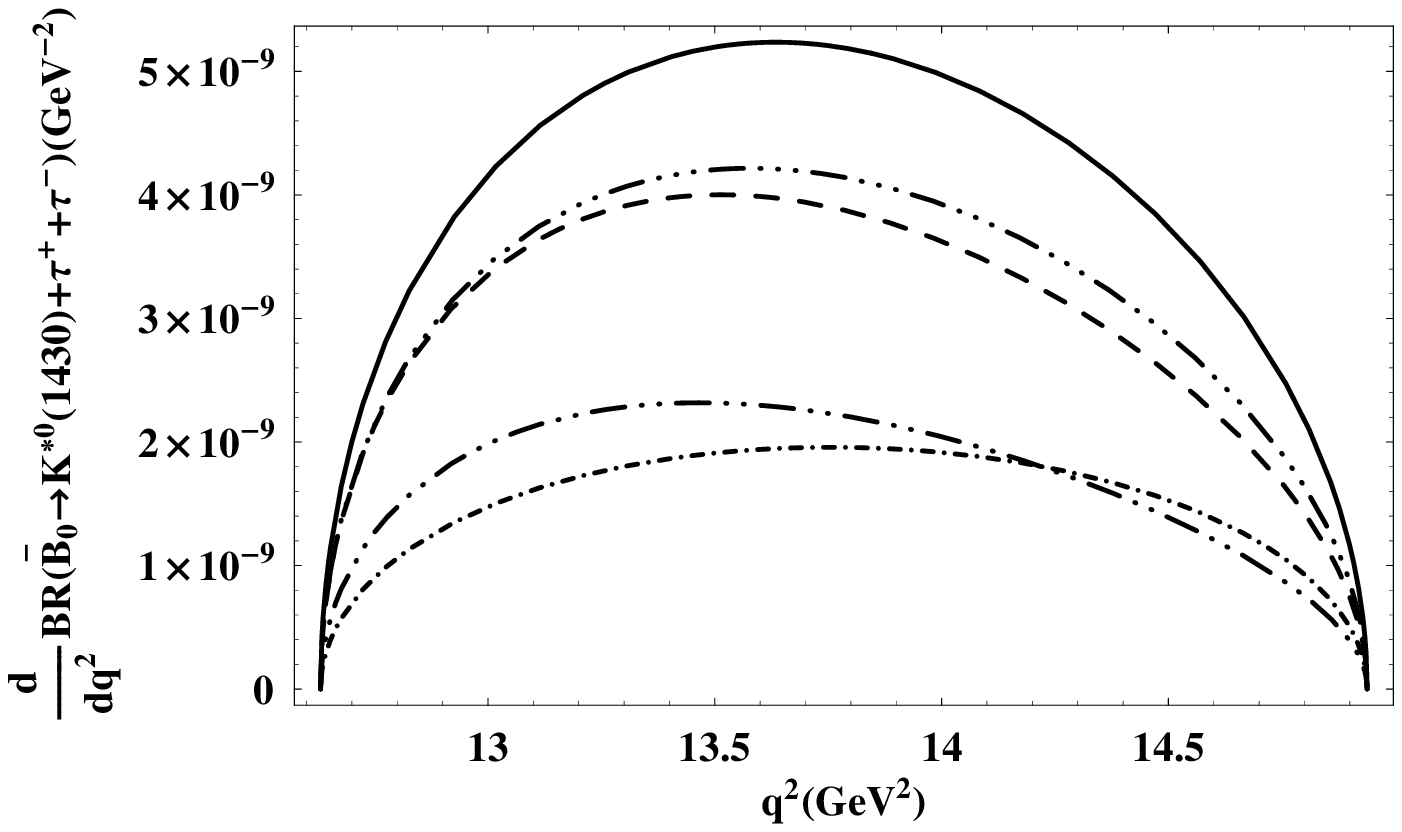}
\put (-350,200){(a)} \put (-100,200){(b)}
\end{tabular}
\end{center}
\caption{The differential width for the $B \to K^{\ast}_0 l^+l^-$ ($%
l=\mu, \tau$) decays as functions of $q^2$. The solid, dashed,
dashed-dot, dashed-double dot and dashed-triple dot line represents
SM, SUSY I, SUSY II, SUSY III and SUSY SO(10) GUT model,
respectively. } \label{decay rate}
\end{figure}

The numerical results for the decay rates, forward-backward
asymmetries and polarization asymmetries of the lepton are presented
in Figs. 1-5. Fig. 1 describes the differential decay rate of
$B\rightarrow K^{\ast}(1430) l^{+}l^{-}$, from which one can see
that the supersymmetric effects are quite distinctive from that of
the SM both in the small and large momentum region. The reason for
the increase of differential decay width in SUSY I model is the
relative change in the sign of $C_{7}^{eff}$; while the large change
in SUSY II model is due to the contribution of the NHBs. As for the
SUSY III and SUSY SO (10) models, the value of the Wilson
coefficients corresponding to NHBs is small and hence one expects
small deviations from SM. Similar effects can also be seen for the
tauon case in Fig. 1b.
\begin{figure}[tbp]
\begin{center}
\begin{tabular}{ccc}
\vspace{-2cm} \includegraphics[scale=0.6]{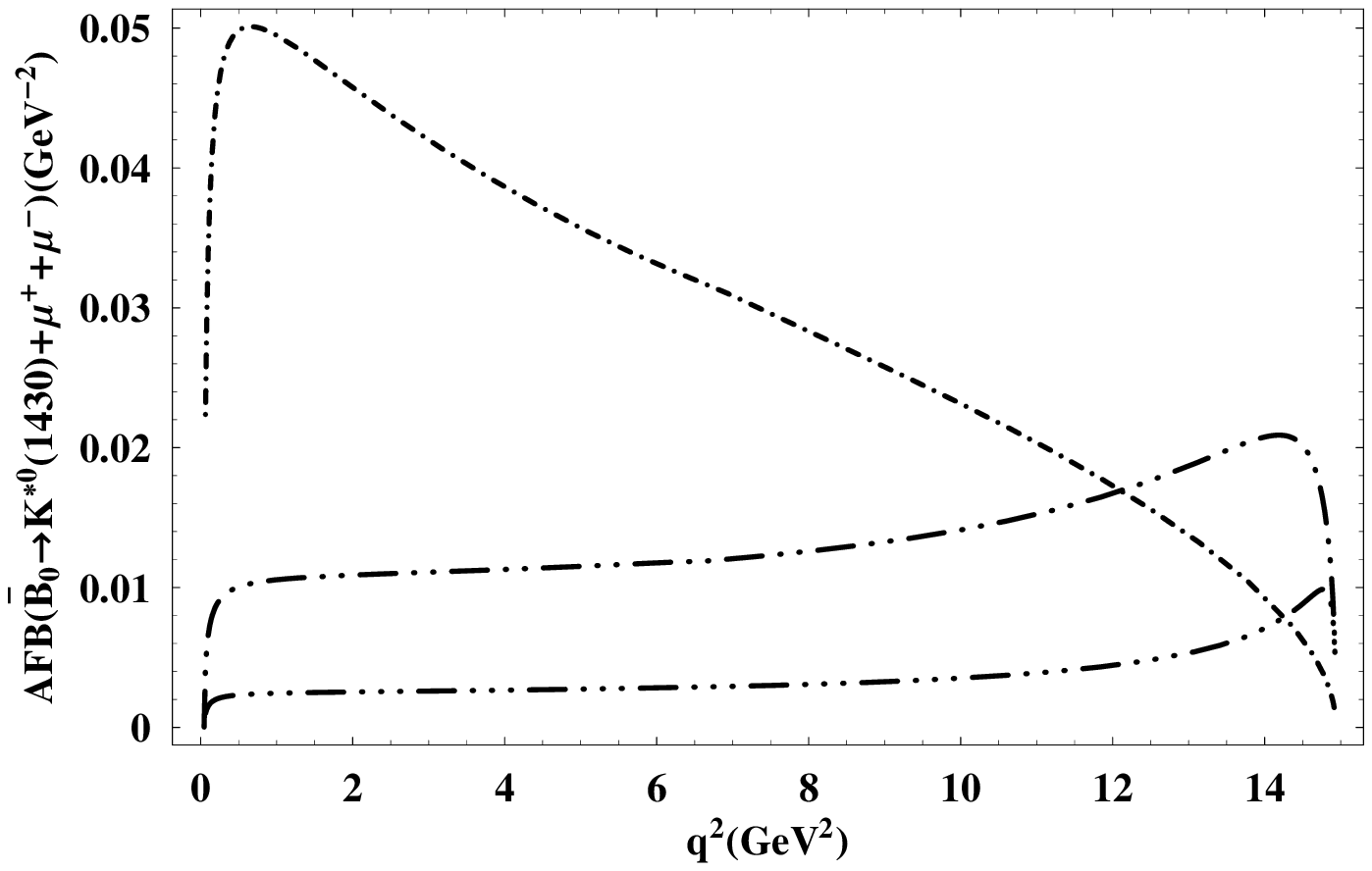} %
\includegraphics[scale=0.6]{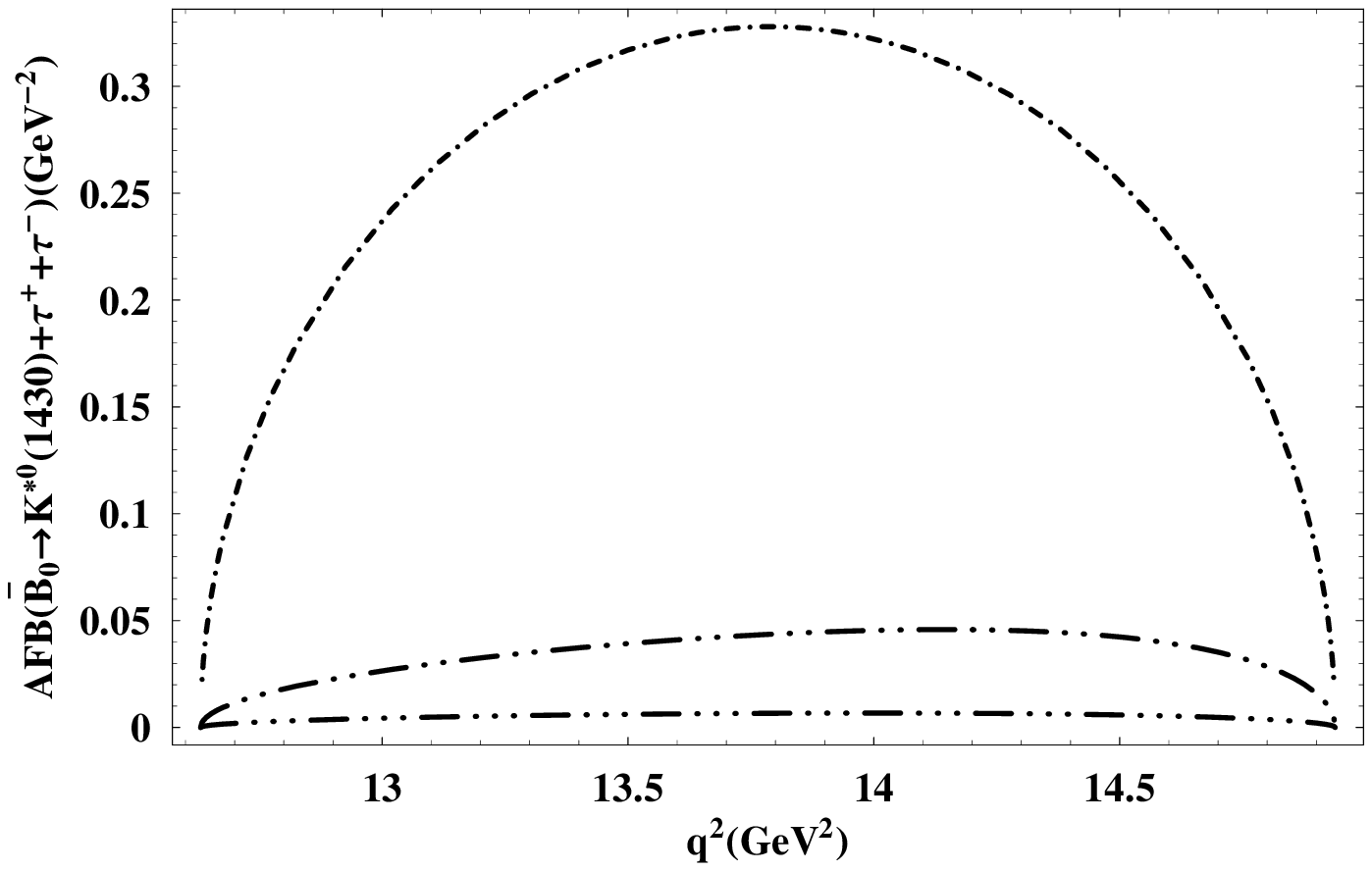} \put (-350,200){(a)} \put (-100,200){(b)}
\end{tabular}
\end{center}
\caption{Forward-backward asymmetry for the $B \to K^{\ast}_0 l^+l^-$ ($%
l=\mu, \tau$) decays as functions of $q^2$. The  dashed-dot,
dashed-double dot and dashed-triple dot line represents SUSY II,
SUSY III and SUSY SO(10) GUT model, respectively.} \label{forward
backward asymmetry}
\end{figure}

\begin{figure}[tbp]
\begin{center}
\begin{tabular}{ccc}
\vspace{-2cm} \includegraphics[scale=0.6]{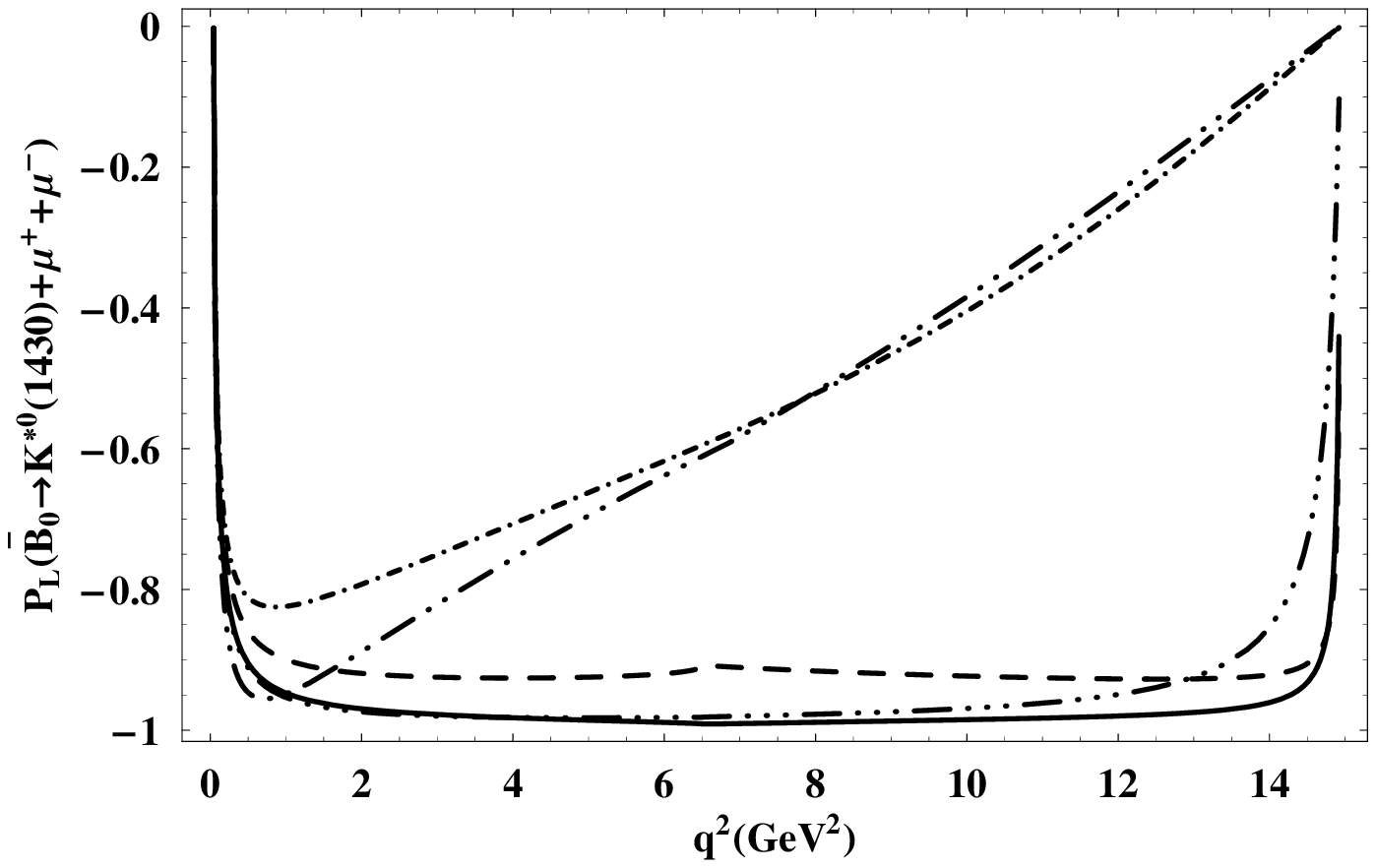} %
\includegraphics[scale=0.6]{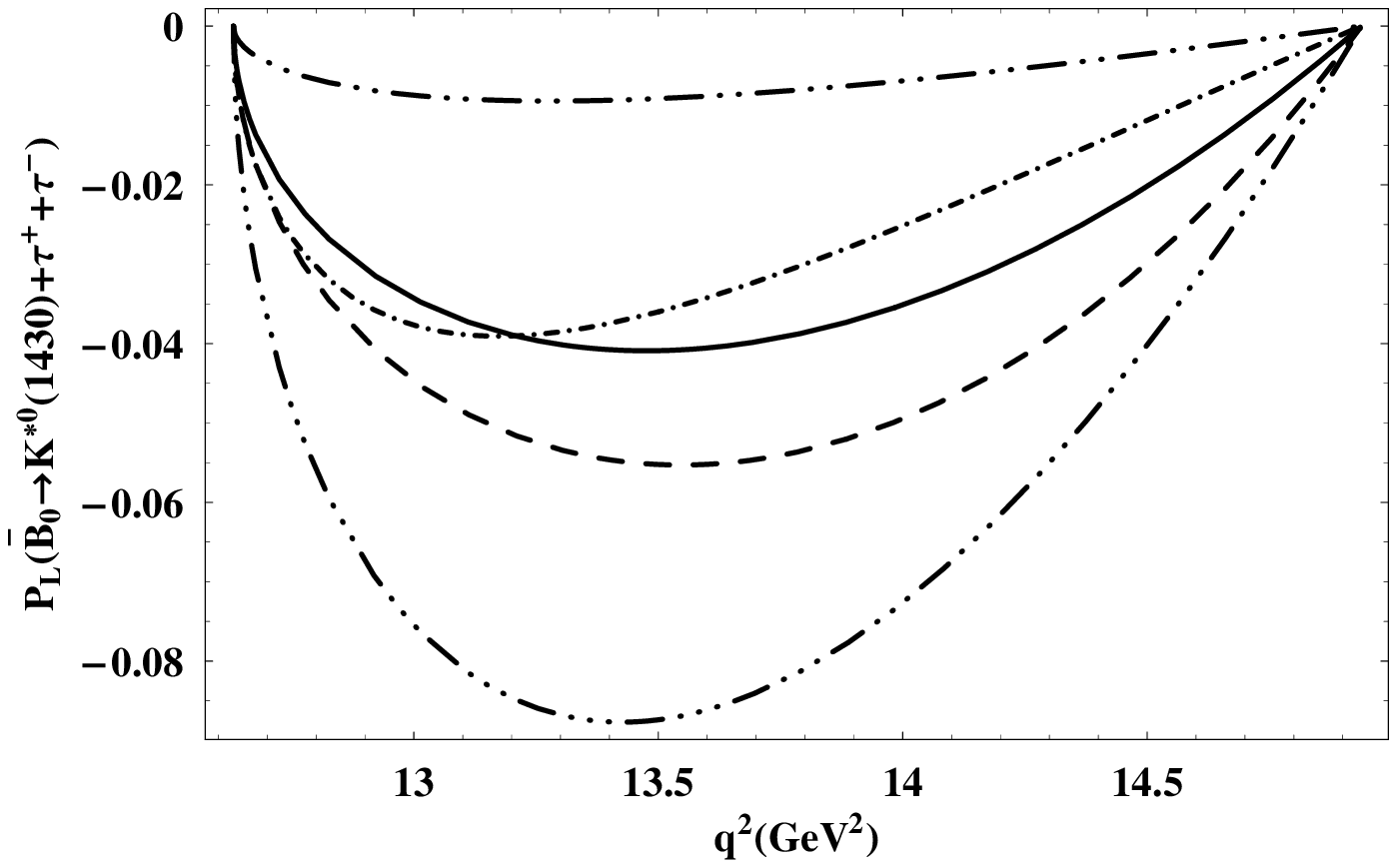}  \put (-350,200){(a)} \put (-100,200){(b)}
\end{tabular}
\end{center}
\caption{Longitudinal lepton polarization asymmetries for the $B \to K^{\ast}_0 l^+l^-$ ($%
l=\mu, \tau$) decays as functions of $q^2$. The solid, dashed,
dashed-dot, dashed-double dot and dashed-triple dot line represents
SM, SUSY I, SUSY II, SUSY III and SUSY SO(10) GUT model,
respectively.} \label{Longitudinal lepton polarization asymmetry}
\end{figure}

In Fig. 2, the forward-backward asymmetries for $B \to K^{\ast}_0 l^+l^-$ ($%
l=\mu, \tau$) are presented. In SM the forward-backward asymmetry is
zero for this decay because there is no scalar operator. However in
SUSY II, SUSY III and SUSY SO(10) model, we have the scalar
operators corresponding to the NHBs, therefore we expect the nonzero
value of the forward-backward asymmetry. This is quite clear from
the Eq. (\ref{FBasymmetry}) where the auxiliary function $D$
corresponds to the contributions from NHBs.  Fig. 2a describes the
forward-backward asymmetry for $B \to K^{\ast}_0 \mu ^{+}\mu ^{-}$.
As the forward-backward asymmetry is proportional to the lepton
mass, therefore for the muons case it is expected to be very small
compared to the tauons case. Thus the maximum value of the
forward-backward asymmetry is $0.05$ in SUSY II model which is hard
to be observed experimentally. However, for $B \to K^{\ast}_0 \tau
^{+}\tau ^{-}$ the maximum value of forward-backward asymmetry is
around $0.35$ in SUSY II model. The number of events required to
observe this asymmetry are around $10^{8}$ or so which are
accessible at large colliders like the LHCb. When the final state
leptons are the tauon pair, the effects of SUSY III and SUSY SO(10)
are still too  small to be measured experimentally.

\begin{figure}[tbp]
\begin{center}
\begin{tabular}{ccc}
\vspace{-2cm} \includegraphics[scale=0.56]{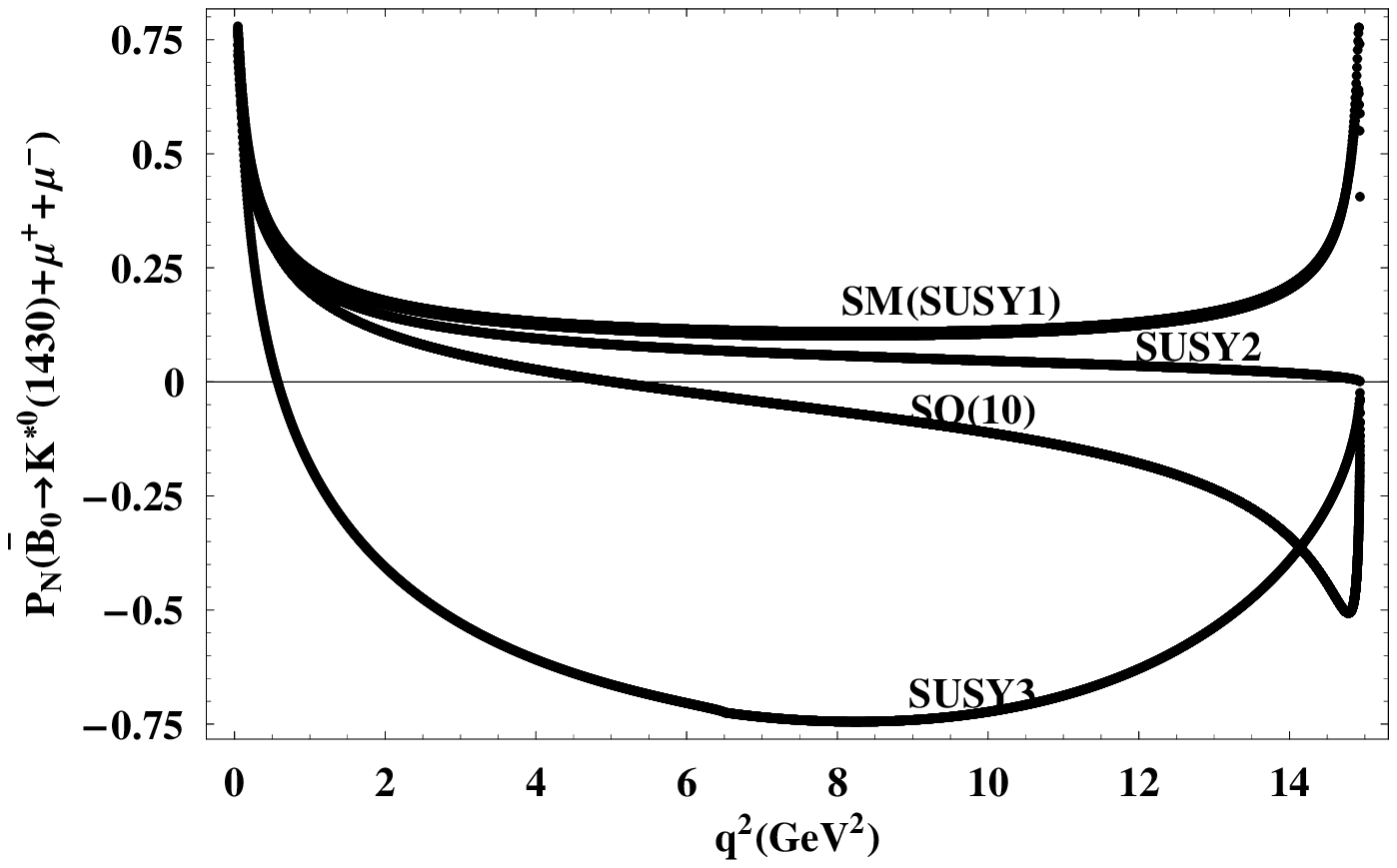} %
\includegraphics[scale=0.56]{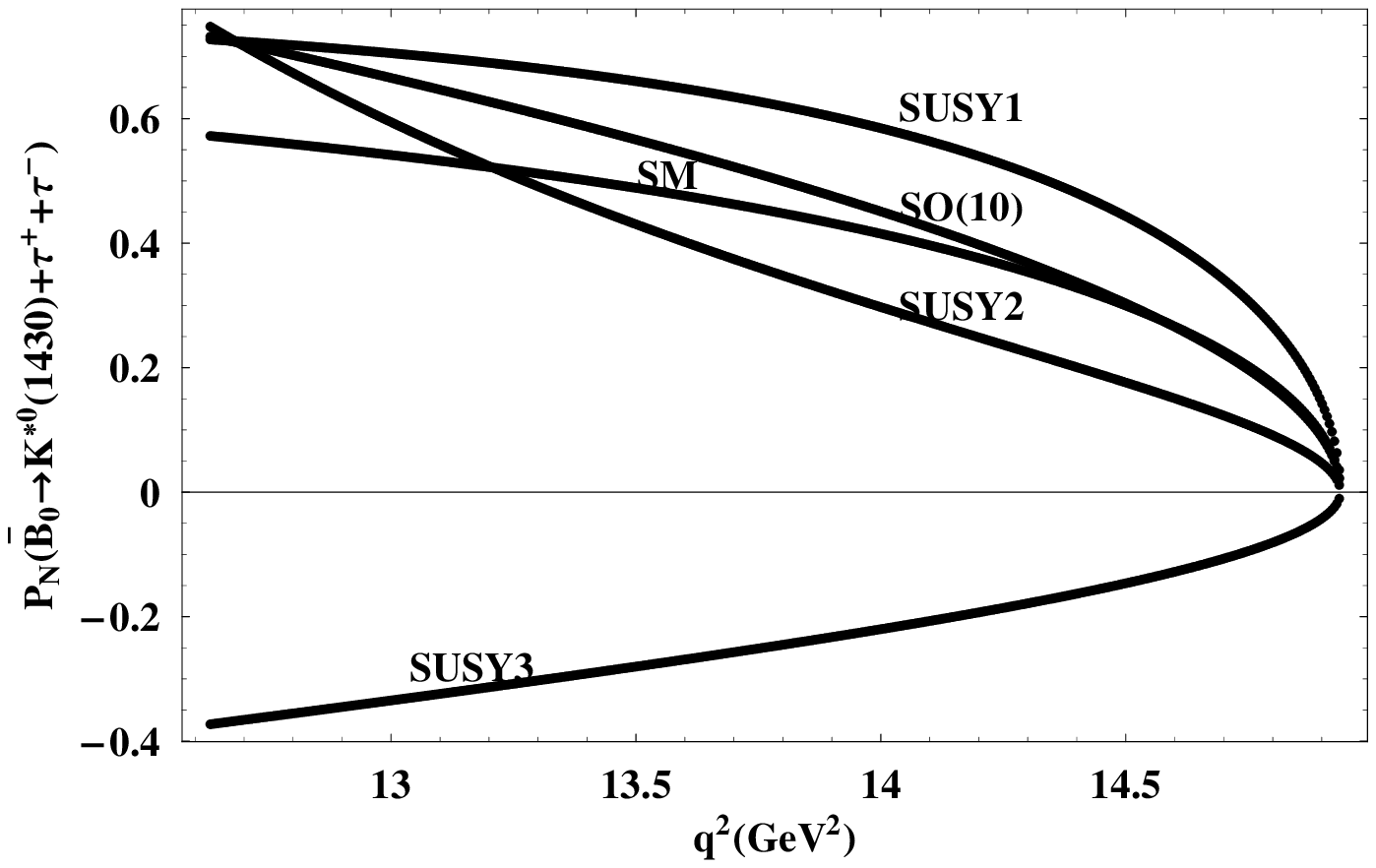} \put (-350,200){(a)} \put
(-100,200){(b)} &  &
\end{tabular}
\end{center}
\caption{Normal lepton polarization asymmetries for the $B \to K^{\ast}_0 l^+l^-$ ($%
l=\mu, \tau$) decays as functions of $q^2$. } \label{Normal lepton
polarization asymmetry}
\end{figure}

Fig. 3(a,b) shows the dependence of longitudinal polarization
asymmetry for the $B \to K^{\ast}_0 l^+l^-$ on the square of
momentum transfer. The value of longitudinal lepton polarization for
muon is around $-1$ in the SM and we have a slight deviation on this
value for SUSY I and SUSY SO(10) model. However, in SUSY II and SUSY
III model the value of longitudinal lepton polarization approaches
to zero in the large momentum transfer region. The reason is that in
SUSY II model we have a large value of the differential decay rate
and this suppresses the value of the polarization in the large
$q^{2}$ region. In SUSY III though the value of the decay rate is
not large, relatively small contribution comes from the Wilson
coefficients $C_{7}^{*eff}C_{10}$. In large $q^{2}$ region, the
longitudinal lepton polarization approaches to zero in all the
models including the SM, because the factor $\lambda
(m_{B}^{2},m_{K^{\ast}_0}^{2},s)$  goes to zero at large value of
transfer momentum. Similar effects can be seen for the final state
tauon but the value for this case is too small to measure
experimentally.

\begin{figure}[tbp]
\begin{center}
\begin{tabular}{ccc}
\vspace{-2cm} \includegraphics[scale=0.58]{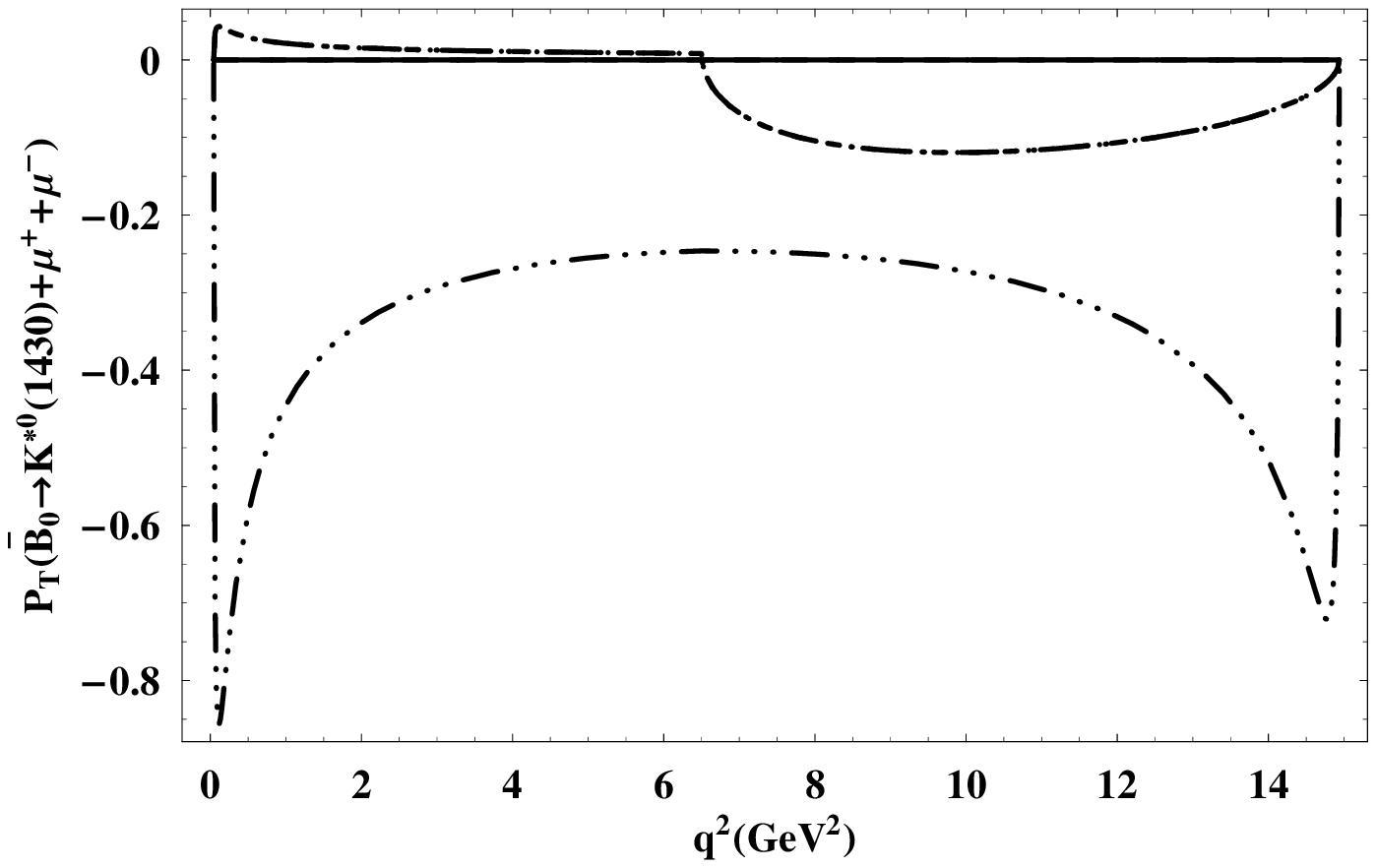} %
\includegraphics[scale=0.58]{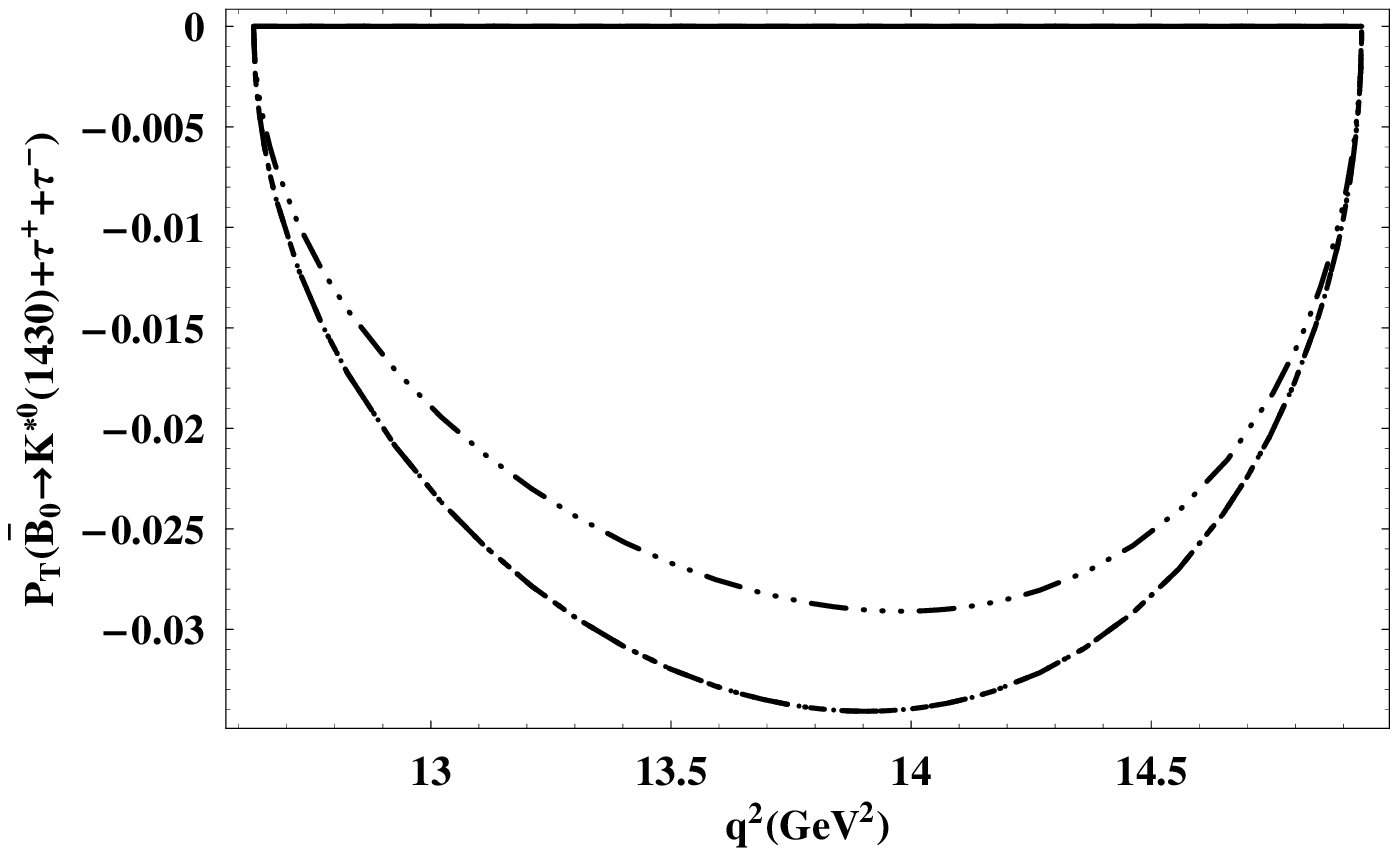} \put (-350,200){(a)} \put
(-100,200){(b)} &  &
\end{tabular}
\end{center}
\caption{Transverse lepton polarization asymmetries for the $B \to
K^{\ast}_0 l^{+}l^{-}$ ($l=\mu, \tau$) decays as functions of $q^2$.
} \label{Transverse lepton polarization asymmetry}
\end{figure}

The dependence of lepton normal polarization asymmetries for $B \to
K^{\ast}_0 l^{+}l^{-}$ on the momentum transfer squared are
presented in Fig. 4. In terms of Eq. (\ref{expression-PN}), one can
observe that this asymmetry is sensitive to the contribution of NHBs
in almost all the supersymmetric models. Fig. 4a shows the normal
lepton polarization for $B \to K^{\ast}_0 \mu^{+} \mu^{-}$. It can
be seen that $P_{N}$ changes its sign in the case of SUSY III model
and this is due to the contribution from NHBs. Now for SUSY II
model, though large contributions from NHBs but it is overshadowed
by the opposite sign of $C_{7}^{*eff}$ and $C_{9}^{eff}$. As the
normal lepton polarization is proportional to the lepton mass, for
$\tau ^{+}\tau ^{-}$ channel, it is expected that one can
distinguish between different SUSY models, which can be seen from
the Fig. 4b. Again due to same reasons as for the muons case, the
normal lepton polarization changes its sign in SUSY III model.

Fig. 5 shows the dependence of transverse polarization asymmetries
for $B \to K^{\ast}_0 l^{+}l^{-}$ on the square of momentum
transfer. From Eq. (\ref{expression-TP}) we can see that it is
proportional to the imaginary part of the Wilson coefficients which
are negligibly small in SM as well as in SUSY I, SUSY II and SUSY
III models. However, complex flavor non-diagonal down-type squark
mass matrix elements of 2nd and 3rd generations are of order one at
GUT scale in SUSY SO(10) model, which induce complex couplings and
Wilson coefficients. As a result, non zero transverse polarization
asymmetries for $B \to K^{\ast}_0 l^{+}l^{-}$ exist in this model.
Now for $\mu ^{+}\mu ^{-}$ channel, the value of transverse
polarization asymmetry is around $-0.3$ in almost all value of
$q^{2}$ except at the end points. Experimentally, to measure
$\left\langle P_{T}\right\rangle $ of a particular decay branching
ratio $\mathcal{B}$ at
the $n\sigma $ level, the required number of events are $N=n^{2}/(\mathcal{B}%
\left\langle P_{T}\right\rangle ^{2})$ and if $\left\langle
P_{T}\right\rangle \sim 0.3$, then the required number of events are almost $%
10^{8}$ for $B$ decays. Since at LHC and BTeV machines, the expected
number of $b\bar{b}$ production events is around $10^{12}$ per year,
so the measurement of transverse polarization asymmetries in the $B
\to K^{\ast}_0 l^{+}l^{-}$  decays could discriminate the SUSY
SO(10) model from the SM and other SUSY models.

\section{Conclusion}

We have carried out the study of invariant mass spectrum,
forward-backward asymmetry, polarization asymmetries of semileptonic
decays $B \rightarrow K^{\ast}_0(1430) l^{+}l^{-}$ ($l=\mu ,\tau $)
 in SUSY theories including SUSY SO(10) GUT model.
Particularly, we analyzed the effects of NHBs to this process and
the main outcomes of this study can be summarized as follows:

\begin{itemize}
\item  The differential decay rates deviate sizably from that of the SM
especially in the large momentum transfer region. These effects are
significant in SUSY II model where the value of the Wilson
coefficients corresponding to the NHBs is large. However, the SUSY
SO(10) effects in differential decay rate of $B \rightarrow
K^{\ast}_0(1430) l^{+}l^{-}$ ($l=\mu ,\tau $) are negligibly small .

\item  The forward-backward asymmetry for the decay $B
\to K^{\ast}_0 l^{+}l^{-}$ is zero in the SM because of the missing
of scalar operators in SM. Hence, the SUSY effects show up  and the
maximum value of the forward-backward asymmetry is around $0.35$ for
$B \to K^{\ast}_0 \tau^{+} \tau^{-}$ in SUSY II model. When the
final state leptons are the tauon pair, the effects of SUSY III and
SUSY SO(10) are still too small to be measured experimentally.

\item  The longitudinal, normal and transverse polarizations of leptons are
calculated in different SUSY models. It is found that the SUSY
effects are very promising which could be measured at future
experiments and shed light on the new physics signal beyond the SM.
The transverse polarization asymmetry is the most interesting
observable to look for the SUSY SO(10) effects where its value is
around $0.3$ in almost all the $q^{2}$ region. It is   measurable at
future experiments like LHC and BTeV machines where a large number
of $\ b \bar {b} $ pairs are expected to be produced.
\end{itemize}

In short, the experimental investigation of observables, like decay
rates, forward-backward asymmetry and lepton polarization
asymmetries
  in $B \rightarrow K^{\ast}_0(1430) l^{+}l^{-}$
($l=\mu ,\tau $) decay will be used to search for the SUSY effects,
in particular the NHBs effect, encoded in the MSSM as well as SUSY
SO(10) models.

\section*{Acknowledgements}

This work is partly supported by National Science Foundation of
China under Grant No.10735080 and 10625525. The author Jamil Aslam
would like to thank C. D. L\"{u} for the kind hospitality in
Institute of High Energy Physics, CAS, where part of this work has
been done.

\end{document}